\documentstyle[12pt]{article}

\makeatletter
\@addtoreset{equation}{section}
\makeatother

\def\theequation{\thesection.\arabic{equation}}

\newcommand{\be}{\begin{equation}}
\newcommand{\ee}{\end{equation}}
\newcommand{\bee}{\begin{eqnarray}}
\newcommand{\eee}{\end{eqnarray}}

\newcommand{\ga}{\alpha}
\newcommand{\gb}{\beta}
\newcommand{\gga}{\gamma}
\newcommand{\gd}{\delta}
\newcommand{\gl}{\lambda}

\newcommand{\gep}{\epsilon}
\newcommand{\gvep}{\varepsilon}
\newcommand{\gs}{\sigma}
\newcommand{\go}{\omega}

\newcommand{\half}{\frac{1}{2}}
\newcommand{\z}{\bar{z}}
\newcommand{\q}{\bar{q}}
\newcommand{\n}{\bar{n}}
\newcommand{\W}{\bar{W}}
\newcommand{\Y}{\bar{Y}}
\newcommand{\bxi}{\bar{\xi}}
\newcommand{\dt}{\partial_{\tau}}

\begin{document}

\thispagestyle{empty}

\begin{flushright}
\vspace{1mm}
FIAN/TD/17--99\\
{June 1999}\\
\end{flushright}
{}~\hfill hep-th/9907020 \\

\vspace{1cm}

\begin{center}
{\large\bf
COHOMOLOGY OF ARBITRARY SPIN CURRENTS \\
IN $AdS_3$}\\
\vglue 1  true cm
\vspace{2cm}
{\bf S.~F.~Prokushkin
\footnote{e-mail: prok@td.lpi.ac.ru}
  and M.~A.~Vasiliev }
\footnote{e-mail: vasiliev@td.lpi.ac.ru}  \\
\vspace{1cm}

I.E.Tamm Department of Theoretical Physics, Lebedev Physics
Institute,\\
Leninsky prospect 53, 117924, Moscow, Russia
%\medskip
\vspace{1.5cm}
\end{center}

\begin{abstract}
We study conserved currents of any
integer or half integer spin built from massless
scalar and spinor fields in $AdS_3$.
2-forms dual to the conserved currents in $AdS_3$ are
shown to be exact in the class of infinite
expansions in higher derivatives of the matter fields
with the coefficients containing inverse powers of the
cosmological constant. This property has no analog in the
flat space and may be related  to the holography
of the AdS spaces. ``Improvements'' to the physical currents are
described as the trivial local current cohomology class.
A complex of spin $s$ currents $(T^s, {\cal D})$ is defined
and the cohomology group $H^1(T^s, {\cal D}) = {\bf C}^{2s+1}$
is found. This paper is an extended version of hep-th/9906149.

\end{abstract}

\newpage

\section{Introduction}\label{intro}

The role of anti-de Sitter (AdS) geometry in the high energy
physics increased greatly due to the Maldacena conjecture \cite{mald}
on the duality between the theory of gravity in the AdS space and
conformal theory on the boundary of the AdS space \cite{adscft,holo}.
The holography hypothesis suggests that the two types of
theories are equivalent. The same time, AdS geometry plays very
important role in the theory of higher spin gauge fields
(for a brief review see \cite{rev}) because interactions of higher
spin gauge fields contain negative powers of the cosmological
constant \cite{FV}. The theory of higher spin gauge fields may be
considered \cite{rev} as a candidate for a most symmetric phase of
string theory.

The group manifold case of $AdS_3$ is special and interesting
in many respects. In particular, the 2d models on the boundary of
$AdS_3$ are conformal \cite{AdS3}. From the higher spin
perspective, a special feature of 3d models is that higher spin
gauge fields are not propagating in analogy with the usual
Chern-Simons gravitational and Yang-Mills fields.
Nevertheless the higher spin gauge symmetries remain
nontrivial, like the gravitational (spin 2) and inner (spin 1)
symmetries.  The higher spin currents can be constructed from the
matter fields of spin 0 and spin 1/2. Their couplings to higher spin
gauge potentials describe interactions of the matter via higher
spin gauge fields.

Schematically, the equations of motion in the gauge field sector
have a form
\be
\label{RJ}
   R = J(C;W) \,,
\ee
where $R=dW-W\wedge W$ denotes all spin $s\geq 1 $ curvatures
built from the higher spin potential $W$, while $C$ denotes
the matter fields (precise definitions are given in the
sect.~\ref{matter}). The 2-form $J(C;W)$ dual to the
3d conserved current vector field obeys the conservation law
\be
\label{DJ}
   DJ(C;W) = 0\,
\ee
as a consequence of the equations of motion in the matter field
sector. $D$ is the covariant derivative of the (infinite-dimensional)
higher spin gauge symmetry algebra \cite{HS, Blen}, i.e.
$\delta R = D\delta W\,,$ where $\delta W$ is an arbitrary variation
of the higher spin gauge potential.
To analyze the problem perturbatively, one fixes a vacuum solution
$W_0$ that solves
\be
\label{R0}
   R_0  =0\,,
\ee
assuming that $W=W_0 + W_1$ while $C$ starts from the
first-order part.  When gravity is included, as is the case in
the higher spin gauge theories, $W_0$ is different from zero and
describes background geometry. In the lowest nontrivial order
one gets from (\ref{RJ})
\be
\label{R1}
    R_1 \equiv D_0 W_1 = J_2 (C^2) \,,
\ee
where $D_0$ is built from $W_0$ and $J_2 (C^2)$ is the part of
$J(C;W_0 )$ bilinear in $C$.
The conservation law (\ref{DJ}) requires
\be
\label{DJ2}
   D_0 J_2(C^2) = 0
\ee
on the free equations of motion of the matter fields.

A nonlinear system of equations of motion describing higher spin
gauge interactions for the spin 0 and spin 1/2 matter fields
in $AdS_3$ in all orders in interactions has been formulated both
for massless \cite{Eq} and massive \cite{PV} matter fields.
An interesting property of the proposed equations discovered
in \cite{PV} is that there exists a flow generating a mapping of
the full nonlinear system to the free one. This mapping is
a nonlinear field redefinition having a form of infinite power
series in higher derivatives of the matter fields and is therefore
generically nonlocal. The coefficients of such expansions contain
inverse powers of the cosmological constant (the higher derivative
of a matter field the more negative power of the cosmological
constant appears) and therefore do not admit a flat limit.
We call such expansions in
higher derivatives pseudolocal to distinguish them from nonlocal
expressions that cannot be represented by power series in higher
derivatives.

Comparison of the results of \cite{PV} with (\ref{R1})
implies that such a field redefinition
exists in a nontrivial model if
\be
\label{JDU}
   J_2 (C^2) =D_0 U (C^2) \,,
\ee
where $U$ is some pseudolocal functional of the matter fields.
The cohomological interpretation with $D_0$ as de Rahm differential
is straightforward because $D_0^2 = R_0 =0$.
Indeed, from (\ref{DJ2}) it follows that the current $J_2 (C^2)$
should be closed on the free equations for matter fields, while
(\ref{JDU}) implies that it is exact in the class of
pseudolocal functionals.

This fact has been already demonstrated for the spin 2 current in
\cite{PV} where we have found a pseudolocal $U$ for the
stress tensor constructed from a massless  scalar field.
In this paper, we generalize this result to the currents of
an arbitrary integer or half integer spin which contain
a minimal possible number of spacetime derivatives.
The analysis of the currents of an arbitrarily high spin is greatly
simplified by a formalism of generating functions developed
in this paper. This formalism is based on the so-called
unfolded formulation \cite{Unf} of the relativistic equations
which allows us to analyze the problem  algebraically, automatically
taking into account the on--mass--shell character of the problem.

Exact currents with local $U$ containing at most a finite
number of derivatives of the matter fields reproduce ``improvements'',
i.e., modifications of the currents which are trivially conserved.
The new result about AdS space established in this
paper is that the true currents can also be treated as ``improvements''
in the class of pseudolocal expansions. This sounds very suggestive
in the context of the holography hypothesis since the corresponding
field redefinitions may result in nontrivial boundary terms.

The paper is designed as follows. In sect.~\ref{matter} we collect
some facts about the equations of motion of the Chern-Simons higher
spin gauge fields and the ``unfolded'' formulation of the equations
of motion for the massless spin 0 and 1/2 matter fields in $AdS_3$.
In sect.~\ref{gener} we propose a formalism of generating functions
to describe differential forms bilinear in derivatives of the matter
fields. Then, in sect.~\ref{complex} we formulate using this method
the AdS on-mass-shell complex, and in sect.~\ref{cohom} we study its
cohomology, the cohomology of currents. In sect.~\ref{flat} we
discuss what happens in the flat limit.

\section{Higher Spin and Matter Fields In $AdS_3$}\label{matter}

The 3d higher spin gauge fields are described \cite{Blen, Eq}
by a spacetime 1-form $W=dx^\mu W_\mu(y,\psi | x)$
depending on the spacetime coordinates $x^\mu$ ($\mu=0,1,2$),
auxiliary commuting spinor variables $y_{\ga}$
(indices $\ga\,,\,\gb\,,\gga\, =1,2$ are lowered and raised by
the symplectic form
$\gep_{\ga\gb}=-\gep_{\gb\ga}$, $\gep_{ 12} =\gep^{12}=1$,
$A^\ga =\gep^{\ga\gb}A_\gb$, $A_\ga = A^\gb \gep_{\gb\ga}$),
and the central involutive element $\psi$, $\psi^2 = 1$,
\be
\label{W}
  W_\mu (y,\psi | x) = \sum_{n=0}^\infty {1\over 2in!} \;
  \left[\go_{\mu,\;\ga(n)} (x) + \gl\psi \;
  h_{\mu,\;\ga(n)} (x) \right] \;
  y^{\ga_1}\ldots y^{\ga_n}  \,.
\ee
A constant parameter $\gl$ is to be identified with
the inverse radius of $AdS_3$.

The higher spin gauge algebra is a Lie superalgebra
built via (anti)commutators from the associative algebra
spanned by the elements of a form (\ref{W}) with a product law
\be
\label{prod}
   (f*g)(y,\psi) = \frac{1}{(2\pi)^2}\int d^2ud^2v \;
   \exp(iu_\ga v^\ga)\; f(y+u,\psi)\; g(y+v,\psi) \,,
\ee
where the integration variables $u_\ga$ and $v_\ga$ are
two-component spinors (in accordance with (\ref{W}) the boson-fermion
parity is identified with the parity in the auxiliary variables $y$).
This product law yields a particular realization of the Weyl algebra,
$[y_\ga,y_\beta]_* = 2i\epsilon_{\ga\beta}$.
The field strength is \cite{HS, Blen}
\be
\label{R}
   R(y,\psi | x) = dW(y,\psi | x) -
   W(y,\psi | x) * \wedge W(y,\psi | x) \,,
\ee
and the equations of motion for the Chern-Simons higher spin gauge
fields with a matter source have a form (\ref{RJ}).

The role of the element $\psi$ is to make the 3d higher spin
superalgebra semisimple  ($hs(2)\oplus hs(2)$ in notation of
\cite{HS}), with simple components singled out by the projectors
$P_\pm = \half (1\pm\psi )$. This is similar to
the $AdS_3$ isometry algebra $o(2,2)\sim sp(2)\oplus sp(2)$.
The latter is identified with
a subalgebra of $hs(2)\oplus hs(2)$ spanned by
\be
\label{AdS}
   L_{\ga\gb}= {1\over 2i} \; y_\ga y_\gb \,,\qquad
   P_{\ga\gb} = {1\over 2i} \; y_\ga y_\gb \psi  \,.
\ee
We therefore identify the $o(2,2)$ components of $W(y,\psi | x)$
(\ref{W}) with the gravitational Lorentz connection 1-form
$\go^{\ga\gb}(x)=dx^\mu\go_{\mu,}{}^{\ga\gb}(x)$ and the dreibein
1-form $h^{\ga\gb}(x)=dx^\mu h_{\mu,}{}^{\ga\gb}(x)$. Since
$AdS_3$ algebra $o(2,2)$ is a proper subalgebra of the d3 higher spin
algebra it is a consistent ansatz to require the vacuum value of
$W(y,\psi | x)$ to be non-zero only in the spin 2 sector. Then
the equation $R_0 = 0$ is equivalent to the $o(2,2)$ zero-curvature
conditions
\be
\label{d omega}
    d\go_{\ga\gb}=\go_{\ga\gga}\wedge\go_\gb{}^\gga+
\gl^2h_{\ga\gga}\wedge h_\gb{}^\gga   \,,
\ee
\be
\label{dh}
    dh_{\ga\gb}=\go_{\ga\gga}\wedge h_\gb{}^\gga+
  \go_{\gb\gga}\wedge h_\ga{}^\gga \,.
\ee
For the metric interpretation, the dreibein $h_{\nu,}{}^{\ga\gb}$
should be non-degenerate, thus admitting the inverse dreibein
$h^{\nu}{}_{,\,\ga\gb}$,
\be
\label{inv d}
   h_{\nu,}{}^{\ga\gb} h^{\nu}{}_{,\,\gga\gd}=
    \frac12\,(\gd^\ga_\gga\gd^\gb_\gd+\gd^\ga_
    \gd\gd^\gb_\gga)\,
\ee
(we use the normalization convention of \cite{Unf}).
Then, (\ref{dh}) reduces to the zero-torsion condition which
expresses Lorentz connection $\go_{\nu,}{}^{\ga\gb}$ via dreibein
$h_{\nu,}{}^{\ga\gb}$, and (\ref{d omega}) implies that
${\cal R}_{\ga\gb}= -\gl^2h_{\ga\gga}\wedge h_\gb{}^\gga$, where
${\cal R}_{\ga\gb}$ is the Riemann tensor 2-form. Therefore,
the equations (\ref{d omega}) and (\ref{dh}) describe $AdS_3$
with radius $\gl^{-1}$, i.e. $AdS_3$ geometry appears via solution
of the vacuum equation (\ref{R0}).

The massless Klein-Gordon and Dirac equations in $AdS_3$ read
\be
\label{K-G D}
  \Box C = \frac32\gl^2 \; C \qquad   %\E
  \mbox{and} \qquad
  h^{\mu}{}_{,\,\ga}{}^\gb \nabla_{\mu}C_\gb = 0
\ee
for the spin 0 boson field $C(x)$ and spin $1\over 2$ fermion field
$C_{\ga}(x)$. Here $\Box =\nabla^{\mu}\nabla_{\mu}$, where
$\nabla_{\mu}$ is the full covariant derivative with
the symmetric Christoffel connection defined via the metric postulate
\be
\label{nabla}
  \nabla_{\mu} h_{\nu,}{}^{\ga\gb}=0  \,.
\ee
The world indices $\mu$, $\nu$ are raised and lowered by the metric
tensor $g_{\mu\nu}=h_{\mu,}{}^{\ga\gb} h_{\nu,}{}_{\ga\gb}$.

The ``unfolded'' formulation \cite{Unf} of the equations
(\ref{K-G D}) in the form of some covariant constancy conditions
is most convenient for the analysis of cohomology of currents.
To this end one introduces an infinite set of symmetric multispinors
$C_{\ga_1\ldots\ga_{n}}$ for all $n\ge 0$.
({}Following to \cite{Cn} we will assume total symmetrization of
indices denoted by the same letter and will use the notation
$C_{\ga(n)}=C_{\ga_1 \dots \ga_n}$ when only a number of indices is
important.) As shown in \cite{Unf}, the infinite chain of equations
\be
\label{chain}
   D^L C_{\ga(n)} = {i\over 2} \left[ h^{\gb\gga}C_{\gb\gga\ga(n)}
    - \gl^2 n(n-1) \; h_{\ga\ga} C_{\ga(n-2)} \right] \,,
\ee
where $D^L$ is the background Lorentz covariant differential,
\be
\label{DL}
   D^L C_{\ga (n)}=dC_{\ga (n)} + n \; \go_{\ga}{}^\gga
       C_{\gga\ga(n-1)} \,,
\ee
is equivalent to the equations (\ref{K-G D}) for the lowest rank
components $C$ and $C_{\ga}$ along with some
constraints expressing highest multispinors via highest
spacetime derivatives of $C$ and $C_{\ga}$ according to
\bee
\label{der}
   C_{\ga (2n)}(x) &=&
   (-2i)^n\; h^{\nu_1}{}_{,\,\ga\ga} h^{\nu_2}{}_{,\,\ga\ga}
   \ldots h^{\nu_n}{}_{,\,\ga\ga}\;
   \nabla_{\nu_1} \nabla_{\nu_2}\ldots \nabla_{\nu_n} \; C(x) \,,
   \nonumber \\
   C_{\ga (2n+1)}(x) &=& (-2i)^n\;
   h^{\nu_1}{}_{,\,\ga\ga} h^{\nu_2}{}_{,\,\ga\ga}\ldots
   h^{\nu_n}{}_{,\,\ga\ga}\;
   \nabla_{\nu_1}\nabla_{\nu_2}\ldots \nabla_{\nu_n}\; C_{\ga}(x) \,,
\eee
where $\nabla_{\mu}$ is a full background derivative obeying
(\ref{nabla}) (for multispinors
$\nabla_{\mu} C_{\ga(n)} = D^L_{\mu} C_{\ga(n)}$).

Following \cite{Unf}, let us introduce the generating function
\be
\label{Cy}
  C(y,\psi | x) = \sum_{n=0}^\infty
  {1\over n!}\; (\gl^{-1}\psi)^{[\frac n2]}  \;
  C_{\ga_1 \ldots \ga_n}(x) \;
  y^{\ga_1}\ldots y^{\ga_n} = \gl^{\half\pi (C)} \;
  \tilde{C}(\gl^{-\half} y, \psi | x)\,,
\ee
where $[n + a] = n$, $\forall n \in Z$ and $0\leq a <1$, and
the boson-fermion parity $\pi (C)=0(1)$ for even (odd) functions
$C(y)$. The equations (\ref{chain}) can be rewritten
in the form \cite{Unf},
\be
\label{DC}
  D^L C(y,\psi) = {i\gl \over 2}\,\psi \; h^{\ga\gb}
  \left[\frac{\partial}{\partial y^\ga }
  \frac{\partial}{\partial y^\gb}
  - y_\ga y_\gb \right] \; C(y,\psi) \,,
\ee
where $D^L = d - \go^{\ga\gb}y_\ga \frac{\partial}{\partial y^\gb}$.

Let us note that the definition (\ref{Cy}) contains inverse powers of
$\gl$ to obtain (\ref{chain}) from (\ref{DC}) or, equivalently,
to have (\ref{der}) with $\gl$ independent coefficients.
Eq.~(\ref{Cy}) is a manifestation of the general property that
the higher derivatives in the theory appear together with the
negative powers of the cosmological constant.

The fields $C_{\ga_1 \ldots \ga_n}$ are identified with all
on-mass-shell nontrivial derivatives of the matter fields according
to (\ref{der}).
The condition that the system is on--mass--shell is encoded in the
fact that the multispinors $C_{\ga_1 \ldots \ga_n}$ are totally
symmetric. This allows us to work with $C_{\ga_1 \ldots \ga_n}$
instead of explicit derivatives of the matter fields.

Consider now a function $F[C_{\ga (n)}(x)]$ of all components of
$C_{\ga_1 \ldots \ga_n}(x)$ at some fixed point $x$. $F$ is
not supposed to contain any derivatives with respect to the spacetime
coordinates $x$ and therefore looks like a local function of matter
fields. One has to be careful however because, when the equations
(\ref{chain}) hold, (\ref{der}) is true. We will therefore call
a function $F[C_{\ga (n)}]$ pseudolocal if it is an infinite
expansion in the field variables $C_{\ga (n)} (x)$ and local if
$F$ is a polynomial with a finite number of nonzero terms.

In terms of the generating functions $C(y,\psi |x)$ this can be
reformulated as follows. Let $F(C|x)$ be some functional
of the generating function $C(y,\psi | x)$ at some fixed point of
spacetime $x$. According to (\ref{der}) its spacetime locality is
equivalent on--mass--shell to the locality in the $y$ space.
Indeed, from (\ref{DC}) it follows that the derivatives in the spinor
variables form in a certain sense a square root of the spacetime
derivatives. (This is also obvious from (\ref{der}).)

The equation (\ref{R1}) for the d3 higher spin system reads
(in the rest of the paper we use the symbol $D$ instead of $D_0$)
\be
\label{Ch-S}
   D W_1(y,\psi | x) = J(C^2) (y,\psi | x)
\ee
with the background AdS covariant differential
\be
\label{D}
   D = D^L - \gl\psi\; h^{\ga\gb} \;
   y_{\ga}\frac{\partial}{\partial y^\gb} = d - (\go^{\ga\gb}
   + \gl\psi\; h^{\ga\gb}) \;
   y_{\ga}\frac{\partial}{\partial y^\gb} \,.
\ee
That  $\go^{\ga\gb}(x)$ and $h^{\ga\gb}(x)$ obey the equations
(\ref{d omega}) and (\ref{dh}) guarantees $D^2 =0$.
Thus, our problem is to study the cohomology of $D$ (\ref{D}).
Clearly, $D$ commutes with the
Euler operator $N=y^\ga \frac{\partial}{\partial y^\ga}$.
Its eigenvalues are identified with spin $s$,
\be
   N = 2(s-1)\,.
\ee
The problem therefore is to be analyzed for different spins
independently.

Conserved currents of an arbitrary integer spin in d4 Minkowski
spacetime were considered in \cite{BBD}. For $d=2$, higher spin
conserved currents were constructed in \cite{BB}. Also,
some currents of spin higher than 2 were recently discussed
in \cite{Ans}.

In the case of $AdS_3$, conserved currents of any integer spin $s\ge 1$
built from two massless scalar fields $C$, $C'$ or two massless spinor
fields $C_\ga$, $C'_\ga$ have a form
\bee
\label{s-scal}
    \hspace*{-1.5cm}
    J^{(s)}_{\mu,\,\ga(2s-2)}(C,C') &=& \sum_{k=0}^{s-2}\;
    {2(-1)^k\over (2k+1)!(2s-2k-3)!}\;
    h_{\mu,}{}^{\gga\gga}\;
    C_{\gga\ga(2k+1)} C'_{\gga\ga(2s-2k-3)} \nonumber\\
    &&\hspace*{-3.5cm}{}+\sum_{k=0}^{s-1} {(-1)^k\over (2k)!(2s-2k-2)!}\;
    h_{\mu,}{}^{\gga\gga}\;
    \left[ C_{\gga\gga\ga(2k)} C'_{\ga(2s-2k-2)}
    - C_{\ga(2k)} C'_{\gga\gga\ga(2s-2k-2)} \right] \,,
\eee
\bee
\label{s-spinor}
    \hspace*{-1cm}
    J^{(s)}_{\mu,\,\ga(2s-2)}(C_\ga,C'_\ga) &=& \sum_{k=0}^{s-1}\;
    {2(-1)^{k+1}\over (2k)!(2s-2k-2)!}\;
    h_{\mu,}{}^{\gga\gga}\;
    C_{\gga\ga(2k)} C'_{\gga\ga(2s-2k-2)} \nonumber\\
    &&\hspace*{-4.3cm}{}+\sum_{k=0}^{s-2}
    {(-1)^k\over (2k+1)!(2s-2k-3)!}\;
    h_{\mu,}{}^{\gga\gga}
    \left[ C_{\gga\gga\ga(2k+1)} C'_{\ga(2s-2k-3)}
    - C_{\ga(2k+1)} C'_{\gga\gga\ga(2s-2k-3)} \right] .
\eee
The supercurrent of any half-integer spin $s\ge {3\over 2}$ built
from massless scalar $C$ and spinor $C'_\ga$ has a form
\bee
\label{supercurrent}
    \hspace*{-1cm}
    J^{(s)}_{\mu,\,\ga(2s-2)}(C,C'_\ga) &=& \sum_{k=0}^{s-{3\over 2}}
    \left\{ {2(-1)^k\over (2k+1)!(2s-2k-3)!}\;
    h_{\mu,}{}^{\gga\gga}\; C_{\gga\ga(2k+1)} C'_{\gga\ga(2s-2k-3)}
    \right. \nonumber\\
    &&\hspace{-3cm}\left.{}+{(-1)^k\over (2k)!(2s-2k-2)!}\;
    h_{\mu,}{}^{\gga\gga}\;
    \left[ C_{\gga\gga\ga(2k)} C'_{\ga(2s-2k-2)}
    - C_{\ga(2k)} C'_{\gga\gga\ga(2s-2k-2)} \right] \right\} \,.
\eee
The lowest spin conserved currents read
\bee
\label{s=1}
   && \hspace*{-2cm}
   J_{\mu}^{(1)}(C,C') = h_{\mu,}{}^{\gga\gga}\;
   (C_{\gga\gga} C' - C C'_{\gga\gga})  \,, \qquad
   J_{\mu}^{(1)}(C_\ga,C'_\ga) = h_{\mu,}{}^{\gga\gga}\;
   C_\gga C'_\gga \,, \\
\label{s=3/2}
   && \hspace*{-2cm}
   J_{\mu,\,\ga}^{({3/2})}(C,C'_\ga) = h_{\mu,}{}^{\gga\gga}\;
   (C_{\gga\gga} C'_\ga - C C'_{\gga\gga\ga}
   + 2 C_{\gga\ga} C'_\gga) \,, \\
\label{s=2}
   && \hspace*{-2cm}
   J_{\mu,\,\ga\ga}^{(2)}(C,C') =
   \half\, h_{\mu,}{}^{\gga\gga}\;
   (C_{\gga\gga} C'_{\ga\ga} - C C'_{\gga\gga\ga\ga}
   - C_{\gga\gga\ga\ga} C' + C_{\ga\ga} C'_{\gga\gga}
   + 4\, C_{\gga\ga} C'_{\gga\ga})  \,.
\eee
These currents are all local because any of them contains a finite
number of terms (i.e., higher derivatives (\ref{der})).
The same expressions remain valid in the flat limit with
$\nabla_\mu \to \partial_\mu$ in (\ref{der}).

\section{Generating Functions}\label{gener}

To analyze the cohomology problem for currents of an arbitrary spin
we first elaborate a technique operating with the generating
functions (\ref{Cy}) rather than with the individual multispinors.

A generic Lorentz covariant spacetime 1-form of spin $s=n/2+1$
bilinear in two different matter fields $C$ and $C^\prime$ and
their on--mass--shell nontrivial derivatives is
\bee
\label{genform}
    \Phi_{\ga (n)}\,(C,C'|x) &=&
    \sum_{k+l=n-2} \sum_{m=0}^\infty a(k,l,m)\;
    h_{\ga\ga} C_{\ga(k)}{}^{\gb(m)}(x)\; C'_{\ga(l)\gb(m)}(x)
    \nonumber \\
    &&\hspace*{-3cm}{}+ \sum_{k+l=n-1} \sum_{m=0}^\infty
    \left[ b_1(k,l,m)\;
    h_\ga{}^\gga C_{\gga\ga(k)}{}^{\gb(m)}(x)\; C'_{\ga(l)\gb(m)}(x)
    \right.\nonumber\\
    &&\hspace*{-3cm}\left.{}+ b_2(k,l,m)\;
    h_\ga{}^\gga C_{\ga(k)}{}^{\gb(m)}(x)\;
    C'_{\gga\ga(l)\gb(m)}(x) \right]   \nonumber \\
    &&\hspace*{-3cm}{}+ \sum_{k+l=n} \sum_{m=0}^\infty
    \left[ e_1(k,l,m)\; h^{\gga\gga} C_{\gga\gga\ga(k)}{}^{\gb(m)}(x)\;
    C'_{\ga(l)\gb(m)}(x)  \right. \nonumber \\
    &&\hspace*{-3cm}\left.{}+ e_2(k,l,m)\;
    h^{\gga\gga} C_{\gga\ga(k)}{}^{\gb(m)}(x)\;
    C'_{\gga\ga(l)\gb(m)}(x)  \right.  \nonumber \\
    &&\hspace*{-3cm}\left.{}+ e_3(k,l,m)\;
    h^{\gga\gga} C_{\ga(k)}{}^{\gb(m)}(x)\;
    C'_{\gga\gga\ga(l)\gb(m)}(x)  \right] \,,
\eee
where $a(k,l,m)$, $b_{1,2}(k,l,m)$, and $e_{1,2,3}(k,l,m)$ are
arbitrary constants and $h_{\ga\ga}$ is the dreibein 1-form.
Introducing
\be
\label{phiy}
   \Phi(y,\psi | x) = \Phi_{\ga_1 \ldots \ga_n}(\psi | x) \;
   y^{\ga_1}\ldots y^{\ga_n}  \,,
\ee
one can equivalently rewrite this formula as
\bee
\label{int1}
  \Phi(y,\psi | x) &=& h_{\ga\ga} \;
    {1\over (2\pi)^2} \oint dr  \oint ds
    \oint \tau^{-2}\,d\tau \int d^2 u \, d^2 v \;
    \exp\left\{{i \over \tau} (u_\gga v^\gga)\right\} \nonumber \\
  &&\hspace*{-2cm} {}\times C(u-ry,\psi | x)\; C'(v+sy,\psi | x)
    \left[ f_1(r,s,\tau)\, y^{\ga} y^{\ga}
    + f_2(r,s,\tau)\, y^{\ga} u^{\ga}
    + f_3(r,s,\tau)\, y^{\ga} v^{\ga} \right. \nonumber \\
  &&\hspace*{-2cm}\left. {}+ f_4(r,s,\tau)\, u^{\ga} u^{\ga}
    + f_5(r,s,\tau)\, u^{\ga} v^{\ga}
    + f_6(r,s,\tau)\, v^{\ga} v^{\ga} \right] \,.
\eee
Here $r$, $s$, and $\tau$ are complex variables, $u_{\ga}$ and
$v_{\ga}$ $(\ga=1,2)$ are spinor variables. The quantities
$f_i(r,s,\tau)$, $i=1, \ldots, 6$ are polynomials in $r^{-1}$
and $s^{-1}$ and formal series in $\tau^{-1}$,
\bee
\label{f1}
    f_1(r,s,\tau) &=& \sum_{0<k,\,l<p, \atop p<\infty} \sum_{m=1}^\infty
    f_1(k,l,m)\; r^{-k} s^{-l} \tau^{-m}   \,, \\
\label{f23}
    f_{2,3}(r,s,\tau) &=& \sum_{0<k,\,l<p, \atop p<\infty}
    \sum_{m=2}^\infty
    f_{2,3}(k,l,m)\;  r^{-k} s^{-l} \tau^{-m}   \,, \\
\label{f456}
    f_{4,5,6}(r,s,\tau) &=& \sum_{0<k,\,l<p, \atop p<\infty}
    \sum_{m=3}^\infty
    f_{4,5,6}(k,l,m)\;  r^{-k} s^{-l} \tau^{-m}   \,.
\eee
The contour integrations are normalized as
$\oint \tau^{-n}d\tau = \gd_n^1$. The Gaussian integrations
with respect to $u_\ga$ and $v_\ga$ should be completed prior
the contour integrations.

Inserting (\ref{Cy}) in the form
\bee
\label{C1}
   &&\hspace*{-1.5cm} C(u-ry,\psi | x) =
   \sum_{n,m=0}^{\infty} {1\over n!m!} \;
   (\gl^{-1}\psi)^{\left[{n+m \over 2}\right]} \;
   C_{\ga(n)\gb(m)}(x) \; u^{\ga_1}\ldots u^{\ga_{n}} \;
   (-r)^m y^{\gb_1}\ldots y^{\gb_{m}} \,, \\
\label{C2}
   &&\hspace*{-1.5cm} C'(v+sy,\psi | x) =
   \sum_{n,m=0}^{\infty} {1\over n!m!} \;
   (\gl^{-1}\psi)^{\left[{n+m \over 2}\right]} \;
   C'_{\ga(n)\gb(m)}(x) \; v^{\ga_1}\ldots v^{\ga_{n}} \;
   s^m y^{\gb_1}\ldots y^{\gb_{m}}
\eee
into (\ref{int1}) and completing elementary integrations
one arrives at (\ref{genform}) with
\bee
\label{a}
   \hspace*{-2cm} a(k,l,m) &=&
    {(-)^{k+m}i^m\over k!\, l!\, m!}
    (\gl^{-1}\psi)^{\left[{k+m\over 2}\right]
    + \left[{l+m\over 2}\right]}\; f_1(k+1,\,l+1,\,m+1) \,, \\
\label{b1}
   \hspace*{-2cm} b_1(k,l,m) &=&
    {(-)^{k+m} i^{m+1} \over k!\, l!\, m!}
    (\gl^{-1}\psi)^{\left[{k+m+1\over 2}\right]
    + \left[{l+m\over 2}\right]}\; f_3(k+1,\,l+1,\,m+2) \,, \\
\label{b2}
   \hspace*{-2cm} b_2(k,l,m) &=&
    {(-)^{k+m+1} i^{m+1} \over k!\, l!\, m!}
    (\gl^{-1}\psi)^{\left[{k+m\over 2}\right]
    + \left[{l+m+1\over 2}\right]}\; f_2(k+1,\,l+1,\,m+2) \,, \\
\label{e1}
   \hspace*{-2cm} e_1(k,l,m) &=&
    {(-)^{k+m+1} i^{m} \over k!\, l!\, m!}
    (\gl^{-1}\psi)^{\left[{k+m\over 2}\right]
    + \left[{l+m\over 2}\right] + 1}\; f_6(k+1,\,l+1,\,m+3) \,, \\
\label{e2}
   \hspace*{-2cm} e_2(k,l,m) &=&
    {(-)^{k+m} i^{m} \over k!\, l!\, m!}
    (\gl^{-1}\psi)^{\left[{k+m+1\over 2}\right]
    + \left[{l+m+1\over 2}\right]}\; f_5(k+1,\,l+1,\,m+3) \,, \\
\label{e3}
   \hspace*{-2cm} e_3(k,l,m) &=&
    {(-)^{k+m+1} i^{m} \over k!\, l!\, m!}
    (\gl^{-1}\psi)^{\left[{k+m\over 2}\right]
    + \left[{l+m\over 2}\right] + 1}\; f_4(k+1,\,l+1,\,m+3) \,.
\eee
Therefore (\ref{int1}) indeed describes a general Lorentz covariant
1-form bilinear in the matter fields.
From these expressions we see that the formula (\ref{int1}) produces
a spacetime local expression if all the coefficients $f_i$ contain
a finite number of terms in (\ref{f1})-(\ref{f456}) and pseudolocal
if some of the expansions in negative powers of $\tau$
are infinite.

In practice, the following representations of rank $n=0,1,2,3$
differential forms $\Phi_n(x)$ are shown below to be most convenient,
\bee
\label{0-form}
    \Phi_0(y,\psi|x) &=& \psi\; {1\over (2\pi)^2}
    \oint {dz\over z} \oint {d\z\over\z}
    \oint {d\tau\over \tau^2} \int d^2 q\,d^2 \q\;
    \exp\left\{-{1 \over 2\tau} (q_\gga\q^\gga)\right\}   \nonumber \\
    &&\hspace*{-3cm}{} \times
    C\left[ \left.\half(q+\q) - {1\over 2i}(z-\z)y\,,\;
    \psi\,\right| \, x \right]
    C'\left[ \left.{1\over 2i}(q-\q) + \half(z+\z)y\,,\;
    \psi\,\right| \,x\right]
    \; E_0(z, \z, \tau) \,, \\
\label{int}
   \Phi_1(y,\psi|x) &=& h_{\ga\ga} \;
    {1\over (2\pi)^2}\oint dz  \oint d\z
    \oint \tau^{-2}\,d\tau \int d^2 q \, d^2 \q\;
    \exp\left\{-{1 \over 2\tau} (q_\gga \q^\gga)\right\} \nonumber \\
  &&\hspace*{-1.5cm}
    \times C\left[ \left.\half(q+\q) - {1\over 2i}(z-\z)y\,,\;
    \psi\,\right| \,x \right]
    C'\left[ \left.{1\over 2i}(q-\q) + \half(z+\z)y\,,\;
    \psi\,\right| \,x \right]   \nonumber \\
  &&\hspace*{-1.5cm} {}\times\left\{R_1(z,\z,\tau)\, y^{\ga} y^{\ga}
    +{1\over 2\tau \z} W_1(z,\z,\tau)\, y^{\ga} \q^{\ga}
    +{1\over 2\tau z} \W_1(z,\z,\tau)\, y^{\ga} q^{\ga}
    \right. \nonumber \\
  &&\hspace*{-1.5cm}\left.
    {}+{1\over 2\tau^2 \z^2} Y_1(z,\z,\tau)\,\q^{\ga} \q^{\ga}
    +{1\over 2\tau^2 z^2} \Y_1(z,\z,\tau)\, q^{\ga} q^{\ga}
    +{1\over 2\tau^2 z\z} V_1(z,\z,\tau)\, q^{\ga} \q^{\ga}
    \right\} \,, \\
\label{2-form}
    \Phi_2(y,\psi|x) &=&
    -{\gl\over 2}\psi\; h_{\ga\gb} \wedge h^{\gb}{}_{\ga} \;
    {1\over (2\pi)^2}\oint dz  \oint d\z
    \oint \tau^{-2}\,d\tau \int d^2 q \, d^2 \q\;
    \exp\left\{-{1 \over 2\tau} (q_\gga \q^\gga)\right\} \nonumber \\
  &&\hspace*{-1.5cm}
    \times C\left[ \left.\half(q+\q) - {1\over 2i}(z-\z)y\,,\;
    \psi\,\right| \,x \right]
    C'\left[ \left.{1\over 2i}(q-\q) + \half(z+\z)y\,,\;
    \psi\,\right| \,x  \right]
    \nonumber \\
  &&\hspace*{-1.5cm} {}\times\left\{R_2(z,\z,\tau)\, y^{\ga} y^{\ga}
    +{1\over 2\tau \z} W_2(z,\z,\tau)\, y^{\ga} \q^{\ga}
    +{1\over 2\tau z} \W_2(z,\z,\tau)\, y^{\ga} q^{\ga}
    \right. \nonumber \\
  &&\hspace*{-1.5cm}\left.
    {}+{1\over 2\tau^2 \z^2} Y_2(z,\z,\tau)\,\q^{\ga} \q^{\ga}
    +{1\over 2\tau^2 z^2} \Y_2(z,\z,\tau)\, q^{\ga} q^{\ga}
    +{1\over 2\tau^2 z\z}
    V_2(z,\z,\tau)\, q^{\ga} \q^{\ga} \right\} \,, \\
\label{3-form}
    \Phi_3(y,\psi|x) &=& -{\gl^2\over 12} \;
    h_{\ga\gb} \wedge h^{\gb}{}_{\gga} \wedge h^{\gga\ga} \;
    {1\over (2\pi)^2}\oint {dz\over z} \oint {d\z\over\z}
    \oint \,{d\tau\over \tau^2} \int d^2 q\,d^2 \q\;
    \exp\left\{-{1 \over 2\tau} (q_\gga\q^\gga)\right\} \nonumber\\
    &&\hspace*{-3cm}{} \times
    C\left[ \left.\half(q+\q) - {1\over 2i}(z-\z)y\,,\;
    \psi\,\right| \,x \right]
    C'\left[ \left.{1\over 2i}(q-\q) + \half(z+\z)y\,,\;
    \psi\,\right| \,x \right] \; E_3(z,\z,\tau)  \,.
\eee
Here the factors of $\psi$, $-{\gl\over 2}\psi$, and
$-{\gl^2\over 12}$ are introduced for future convenience.
It is not hard to see that the expressions
(\ref{0-form})-(\ref{3-form}) reproduce arbitrary Lorentz
covariant forms bilinear in the matter fields and their on-mass-shell
nontrivial derivatives.

Let $n$, $\n$, and $n_\tau$ be the following operators,
\be
\label{nn}
   n = z\, {\partial\over \partial z} \,,\qquad
   \n = \z\, {\partial\over \partial \z} \,,\qquad
   n_\tau = \tau\, {\partial\over \partial\tau}
\ee
(using the same notations for their eigenvalues).
The quantities $R_{1,2}(z,\z,\tau)$, $W_{1,2}(z,\z,\tau)$,
$\W_{1,2}(z,\z,\tau)$, $Y_{1,2}(z,\z,\tau)$,
$\Y_{1,2}(z,\z,\tau)$, $V_{1,2}(z,\z,\tau)$, and
$E_{0,3}(z,\z,\tau)$ give a non-zero contribution to (\ref{0-form}),
(\ref{int}), (\ref{2-form}), and (\ref{3-form}) when $n$, $\n$, and
$n_\tau$ satisfy the following restrictions:
\be
\label{nonzero}
\begin{array}{|l|ccr|ccr|ccr|}
\hline
\vphantom{\displaystyle{A\atop B}}
R_1,\, R_2\hspace{1cm}
&\quad n&\le &-1 \quad &\quad\n &\le &-1 \quad
&\quad n_\tau &\le &-1 \quad\\
\hline
\vphantom{\displaystyle{A\atop B}}
W_1,\, W_2\hspace{1cm}
&\quad n&\le &-1 \quad &\quad \n &\le & 0 \quad
&\quad n_\tau &\le &-1 \quad\\
\hline
\vphantom{\displaystyle{A\atop B}}
\W_1,\, \W_2\hspace{1cm}
&\quad n&\le & 0 \quad &\quad\n &\le &-1 \quad
&\quad n_\tau &\le &-1 \quad\\
\hline
\vphantom{\displaystyle{A\atop B}}
Y_1\,, Y_2\hspace{1cm}
&\quad n&\le &-1 \quad &\quad\n &\le & 1 \quad
&\quad n_\tau &\le &-1 \quad\\
\hline
\vphantom{\displaystyle{A\atop B}}
\Y_1\,, \Y_2\hspace{1cm}
&\quad n&\le & 1 \quad &\quad\n &\le &-1 \quad
&\quad n_\tau &\le &-1 \quad\\
\hline
\vphantom{\displaystyle{A\atop B}}
V_1,\, V_2,\, E_0,\, E_3
&\quad n&\le & 0 \quad &\quad \n &\le & 0 \quad
&\quad n_\tau &\le &-1 \quad\\
\hline
\end{array}
\ee
Beyond these regions, the coefficients do not contribute
and therefore their values can be fixed arbitrarily.
In particular, the quantities $R_{1,2}$, $W_{1,2}$, $\W_{1,2}$, ...
are defined modulo arbitrary polynomials in $\tau$
\be
\label{arb}
    P(\tau) = \sum_{k=0}^{k_0} P_k \tau^k\,.
\ee

The formulae considered in this section make sense for an arbitrary
dimension of spinors (discarding the question of the completeness of
the basis forms like $h_{\ga\gb}$ and
$h_{\ga\gb} \wedge h^{\gb}{}_{\ga}$). For the two-component spinors
there exist additional equivalence relationships due to the fact that
antisymmetrization over any three two-component spinor indices
gives zero. This is expressed by the identity
\be
\label{tF}
   a_\ga(b_\gb c^\gb)
   + b_\ga(c_\gb a^\gb) + c_\ga(a_\gb b^\gb) = 0  \,
\ee
valid for any three commuting two-component spinors $a_\ga$, $b_\ga$,
and $c_\ga$. As a result, the forms discussed so far are not all
independent. The ambiguity in adding any terms which vanish as
a consequence of (\ref{tF}) can be expressed in a form of some
equivalence (gauge)
transformations of the coefficients in (\ref{int1}), (\ref{int}),
and (\ref{2-form}). We call these equivalence transformations
Fierz transformations.

To derive a functional form of a general Fierz transformation
it is convenient to rewrite (\ref{int}) as
\bee
\label{int2}
   \Phi_1(y,\psi|x) &=& h_{\ga\ga} \; {1\over (2\pi)^2}
    \oint dz \oint d\z \oint dt \int d^2 q \, d^2 \q\;
    \nonumber\\
    &&\hspace*{-2cm}{}\times
    \exp\left\{-{t \over 2}(q_\gga \q^\gga)
    - {i\over 2}  z (\q_\gga y^\gga)
    + {i\over 2} \z (y_\gga q^\gga)     \right\}
    C\left(\left.{q+\q \over 2}\,,\;\psi\right| x\right)
    C'\left(\left.{q-\q \over 2i}\,,\;\psi\right| x \right)
    \nonumber\\
    &&\hspace*{-2cm}{}\times\left\{f'_1(z,\z,t)\, y^{\ga} y^{\ga}
    + f'_2(z,\z,t)\, y^{\ga} \q^{\ga}
    + f'_3(z,\z,t)\, y^{\ga} q^{\ga} \right. \nonumber \\
    &&\hspace*{-2cm} \left.{} + f'_4(z,\z,t)\,\q^{\ga} \q^{\ga}
    + f'_5(z,\z,t)\, q^{\ga} \q^{\ga}
    + f'_6(z,\z,t)\, q^{\ga} q^{\ga}  \right\} \,,
\eee
Using the partial integrations w.r.t. $t$, $z$, and $\z$
in (\ref{int2}) one finds that the transformations,
\bee
\label{F}
    \delta f'_1 = \partial_t \gep \,,\qquad
    \delta f'_2 = i \partial_{\z} \gep \,,\qquad
    \delta f'_3 = -i\partial_{z} \gep \,, \nonumber \\
    \delta f'_3 = \partial_t \eta \,,\qquad
    \delta f'_5 = i \partial_{\z} \eta \,,\qquad
    \delta f'_6 = -i\partial_{z} \eta \,, \nonumber \\
    \delta f'_2 = \partial_t \phi \,,\qquad
    \delta f'_4 = i\partial_{\z} \phi \,,\qquad
    \delta f'_5 = -i\partial_{z} \phi \,,
\eee
with arbitrary parameters $\gep=\gep(z,\z,t)$, $\eta=\eta(z,\z,t)$,
and $\phi=\phi(z,\z,t)$ describe all possible Fierz transformations
of the 1-form (\ref{int2}). From here one derives a form of the Fierz
transformations in the representations (\ref{int}) and
(\ref{2-form}),
\bee
\label{delta R}
   \delta R_{1,2} &=& -\dt \chi_{1,2} \,,    \\
\label{delta W}
   \delta W_{1,2} &=& -\dt  \xi_{1,2} + 2i\n \chi_{1,2} \,,  \\
\label{delta bW}
   \delta \W_{1,2} &=& -\dt \bxi_{1,2} - 2in \chi_{1,2}  \,, \\
\label{delta V}
   \delta V_{1,2} &=& -i(n\xi_{1,2} - \n\bxi_{1,2})  \,, \\
\label{delta Y}
   \delta Y_{1,2} &=& i(\n-1)\xi_{1,2}  \,,   \\
\label{delta bY}
   \delta \Y_{1,2} &=& -i(n-1)\bxi_{1,2}  \,,
\eee
with arbitrary parameters $\chi_{1,2}(z,\z,\tau)$,
$\xi_{1,2}(z,\z,\tau)$, and $\bxi_{1,2}(z,\z,\tau)$.

Let us mention that the relation
\bee
\label{hh}
   h_{\ga\gb}\wedge h_{\gga\gd} = {1\over 4}
   \left(\gvep_{\ga\gga}\; h_{\gb\gl} \wedge h^{\gl}{}_{\gd}
   + \gvep_{\gb\gga}\; h_{\ga\gl} \wedge h^{\gl}{}_{\gd}
   + \gvep_{\gb\gd}\; h_{\ga\gl} \wedge h^{\gl}{}_{\gga}
   + \gvep_{\ga\gd}\; h_{\gb\gl} \wedge h^{\gl}{}_{\gga}  \right) \,,
\eee
which allows one to use the representation (\ref{2-form})
for a 2-form, is itself a consequence of (\ref{tF}).

\section{On-Mass-Shell Current Complex}\label{complex}

In this section we study the on--mass-shell action of the operator
$D$ (\ref{D}) on the differential forms defined in sect.~\ref{gener}.
The advantage of the formulation of the dynamical equations
in the unfolded form (\ref{DC}) is that it expresses the (exterior)
spacetime derivative of $C$ via some operators acting in
the auxiliary spinor space.
As a result, on--mass--shell action of $D$ reduces to some mapping
${\cal D}$ acting on the coefficients in the formulae
(\ref{0-form})-(\ref{3-form}).

Let us consider the example of a 0-form. Using the Leibnitz rule for
$D^L$ and taking into account the equations of motion (\ref{DC}) and
the zero torsion condition $D^L h_{\ga\ga} = 0$ (\ref{dh}), one gets
\bee
\label{D phi0}
  D \Phi_0(y,\psi|x) &=& \left( D^L - \gl \psi h^{\ga\gb}
    y_\ga \frac{\partial}{\partial y^\gb } \right) \Phi_0(y,\psi|x)
    \nonumber \\
  && \hspace{-2cm}{} = {i\gl\over 2} \; h^{\ga\ga} \;
    {1\over (2\pi)^2} \oint {dz\over z} \oint {d\z\over\z}
    \oint {d\tau\over \tau^2} \int d^2 q\,d^2 \q\;
    \exp\left\{-{1 \over 2\tau} (q_\gga\q^\gga)\right\}
    E_0(z,\bar{z},\tau)
    \nonumber \\
  &&\hspace{-2cm} {}\times\left[ 4\; {\partial\over \partial q^\ga}
    {\partial\over \partial \q^\ga} - q_\ga \q_\ga
    +iy_\ga(\z q_\ga - z \q_\ga) - z\z\; y_\ga y_\ga
    -2 y_\ga \left( z\;{\partial\over \partial q^\ga}
    - \z\;{\partial\over \partial \q^\ga}\right) \right] \nonumber \\
  &&\hspace{-2cm} {}\times
    \left\{ C\left[ \left.\half(q+\q)
    - {1\over 2i}(z-\z)y\,,\;\psi\right|x \right]
    C'\left[ \left.{1\over 2i}(q-\q) + \half(z+\z)y
    \,,\;\psi\right|x \right] \right\} \,.
\eee
Completing the partial integration w.r.t. $q$ and $\q$
one arrives at the 1-form $\hat{\Phi}_1 = D \Phi_0$ with
the coefficients
$R_1^{\cal D} (E_0)$, $W_1^{\cal D} (E_0)$, $\W_1^{\cal D}(E_0)$,
$Y_1^{\cal D}(E_0)$, $\Y_1^{\cal D}(E_0)$, and $V_1^{\cal D}(E_0)$
of the form
\bee
\label{^phi1}
  R_1^{\cal D} (E_0) &=& - {i\gl\over 2} E_0(z,\z,\tau) \,,
  \nonumber \\
  W_1^{\cal D} (E_0) &=& i\gl(1-i\tau) E_0(z,\z,\tau) \,,
  \nonumber \\
  \W_1^{\cal D} (E_0) &=& i\gl(1+i\tau) E_0(z,\z,\tau) \,,
  \nonumber \\
  V_1^{\cal D} (E_0) &=& -i\gl(1+\tau^2) E_0(z,\z,\tau) \,,
  \nonumber \\
  Y_1^{\cal D} (E_0) &=& \Y_1^{\cal D} (E_0) = 0 \,.
\eee
Analogously one derives the mapping
$\left.{\cal D}:\; \Phi_i(y)\to \hat{\Phi}_{i+1}(y)
= D \Phi_i(y)\right|_{on-shell}$,
$i=1,2$ on the coefficients of the differential forms (\ref{int}),
(\ref{2-form}), (\ref{3-form}),
\be
\label{calD}
  {\cal D}\; \{R_1, W_1, ... \} =
  \{R_2^{\cal D}, W_2^{\cal D}, ... \}\,,
  \qquad {\cal D}\; \{R_2, W_2, ... \} = E_3^{\cal D}  \,,
\ee
with
\bee
\label{^R}
   R_2^{\cal D} &=&
   -(1-i\tau)nR_1 - (1+i\tau)\n R_1 + 2R_1
   +{i\over 4}(1+i\tau)\dt W_1
   -{i\over 4}(1-i\tau)\dt \W_1   \nonumber \\
   & & {}+{1\over 4}(W_1+\W_1)-\half(nW_1+\n\W_1) \,,  \\
\label{^W}
   W_2^{\cal D} &=& -{i\over 2}\dt [(1+\tau^2)W_1]
   +{3\over 2}(1+i\tau)W_1 + 2(1+\tau^2)\n R_1
   +\half (1-i\tau)\n\W_1 + (1-2n)Y_1   \nonumber \\
   & & {}-\half (1+i\tau)(\n-1)W_1
   +i(1+i\tau)\dt Y_1+\left({3\over 2}-\n \right) V_1
   -{i\over 2}(1-i\tau)\dt V_1  \,, \\
\label{^bW}
   \W_2^{\cal D} &=& {i\over 2}\dt [(1+\tau^2)\W_1]
   +{3\over 2}(1-i\tau)\W_1 + 2(1+\tau^2)nR_1
   +\half (1+i\tau)nW_1 + (1-2\n)\Y_1  \nonumber \\
   & & {}-\half (1-i\tau)(n-1)\W_1
   -i(1-i\tau)\dt \Y_1 + \left({3\over 2}-n\right) V_1
   +{i\over 2}(1+i\tau)\dt V_1 \,, \\
\label{^V}
   V_2^{\cal D} &=& \half (1+\tau^2)(nW_1+\n\W_1)
   +\half (1+i\tau)(\n-1) V_1
   +\half (1-i\tau)(n-1)V_1 + V_1  \nonumber \\
   & & {}+(1+i\tau) nY_1 +(1-i\tau)\n\Y_1  \,, \\
\label{^Y}
   Y_2^{\cal D} &=& \half (1+\tau^2)(\n-1) W_1
   -i(1+\tau^2)\dt Y_1
   +(1+i\tau)Y_1 + (1-i\tau) nY_1  \nonumber \\
   & & {}+\half (1-i\tau)(\n-1) V_1  \,,  \\
\label{^bY}
   \Y_2^{\cal D} &=& \half (1+\tau^2)(n-1)\W_1
   +i(1+\tau^2)\dt \Y_1
   +(1-i\tau)\Y_1 + (1+i\tau)\n\Y_1  \nonumber \\
   & & {}+\half (1+i\tau)(n-1) V_1  \,,
\eee
and
\bee
\label{^G}
   \hspace*{-1.6cm} E_3^{\cal D} &=&   4i n\n (1+\tau^2) R_2\,
    + in\n (W_2+\W_2) - 3\tau (nW_2-\n\W_2) - \tau n\n (W_2-\W_2)
    \nonumber\\
   \hspace*{-2.4cm}&& {} + \dt [(1+\tau^2)(nW_2-\n\W_2)]
    + 2\dt (nY_2-\n\Y_2) - (n-\n)\;\dt  V_2 \nonumber\\
   \hspace*{-2.4cm}&& {} + 2i(n\n-n-\n+1) V_2
    + i(n+\n-2)\;\tau\dt  V_2 + i(1+\tau^2)\;\dt\dt V_2 \nonumber\\
   \hspace*{-2.4cm}&& {} - 3i (nW_2+\n\W_2) + (i+\tau)\; n(n+1) W_2
    + (i-\tau)\;\n(\n+1) \W_2   \nonumber\\
   \hspace*{-2.4cm}&& {} - 4i (nY_2+\n\Y_2) + 2i \;
   \tau\dt  (nY_2+\n\Y_2) + 2i [n(n+1) Y_2 + \n(\n+1) \Y_2]  \,.
\eee
As expected, ${\cal D}^2 = 0$ and therefore the mapping ${\cal D}$
defines a complex $(T, {\cal D})$ with
\be
\label{T}
   T = \bigoplus\limits_{i=0,1,2,3} T_i \,,
   \qquad T_{0,3} = \{E_{0,3} \} \,, \;
   T_{1,2} = \{R_{1,2}, W_{1,2}, \W_{1,2}, V_{1,2},
   Y_{1,2}, \Y_{1,2}\} \,.
\ee
The reformulation of the problem in terms of $(T, {\cal D})$
effectively accounts the fact that the fields are on--mass--shell.
We identify the cohomology of currents with the cohomology of
the operator ${\cal D}$ acting on the space $T$ (\ref{T}).

The remarkable property of the mapping ${\cal D}$ is that
it contains $z$, $\z$, ${\partial\over\partial z}$, and
${\partial\over\partial \z}$ only via $n$ and $\n$ (\ref{nn}),
thus implying the separation of variables: the differential
${\cal D}$ leaves invariant eigensubspaces of $n$ and $\n$.
As a result one can consider separately functions $R_{1,2}$,
$W_{1,2}$, ... with equal values $n$ and $\n$.
In fact, this is the main reason for using the particular
representation (\ref{0-form})-(\ref{3-form}). Needless to say
that this property greatly simplifies the study of the cohomology
of currents, reducing it to the analysis of functions of a single
variable $\tau$ with two integer parameters $n$ and $\bar{n}$.
The fact of the existence of such a separation of variables is
a consequence of the form of the matter field equations (\ref{DC}).

As expected, the system (\ref{^R})-(\ref{^bY}) is consistent
with the Fierz transformations (\ref{delta R})-(\ref{delta bY}).
Namely, any Fierz transformation of the quantities $R_1$, $W_1$, ...
leads to the Fierz transformation of the quantities
$R_2^{\cal D}$, $W_2^{\cal D}$, ... with the parameters
\bee
\label{FF1}
   \hspace*{-2cm}
   \chi_2\,(\chi_1, \xi_1, \bxi_1) &=& -\half \left[(n+\n)
   - i\tau(n-\n) - 4 \right]\; \chi_1 \nonumber\\
   &&{}+ {1\over 4} \left[i\dt (\xi_1 - \bxi_1) + (1 - \tau\dt)
   (\xi_1 + \bxi_1) - 2 (n\xi_1 + \n\bxi_1) \right]  \,, \\
\label{FF2}
   \hspace*{-2cm}
   \xi_2\,(\chi_1, \xi_1, \bxi_1) &=& (1+\tau^2)\; \n\chi_1
   + \half (1-i\tau)\; \n\bxi_1  \nonumber\\
   &&{}+ \half\left[-i(1+\tau^2)\; \dt\xi_1 + (n+2)\; \xi_1
   - i\tau (n-2)\; \xi_1 \right]  \,, \\
\label{FF3}
   \hspace*{-2cm}
   \bxi_2\,(\chi_1, \xi_1, \bxi_1) &=& (1+\tau^2)\; n\chi_1
   + \half (1+i\tau)\; n\xi_1 \nonumber\\
   &&{}+ \half\left[i(1+\tau^2)\; \dt\bxi_1 + (\n+2)\; \bxi_1
   + i\tau (\n-2)\; \bxi_1 \right]  \,,
\eee
and any Fierz transformation of $R_2$, $W_2$, ... does not affect
the parameter $E_3^{\cal D}$ (\ref{^G}).

Of course the formulae (\ref{^R})-(\ref{^bY}) are consistent with
the ambiguity in adding trivial terms (\ref{arb}) to the quantities
$R_1$, $W_1$, ... in the sense that this transformation leads to
the analogous transformation of the quantities
$R_2^{\cal D}$, $W_2^{\cal D}$, ... , which does not affect
the 2-form $\hat{\Phi}_2(y)$.

\section{Cohomology of Currents}\label{cohom}

Following \cite{BBD} we study the currents containing
the minimal possible number of spacetime derivatives for a given
spin $s$. {}From (\ref{der}) it is clear that this is the case
if the number of the contracted indices $\gb$ in $(\ref{genform})$
is zero. Since the number of contractions is $-(n_\tau+1)$
(see sect.~\ref{gener}) we consider 2-forms $\Phi_2^{n,\n}$
with $n_\tau = -1$. Thus we set in (\ref{2-form})
\bee
\label{a:t}
   &&\hspace*{-1cm} R_2=\ga_R(n,\n)\; z^n \z^{\n} \tau^{-1}\,,
   \quad
   W_2=\ga_W(n,\n)\; z^n \z^{\n} \tau^{-1}\,, \quad
  \W_2=\ga_{\W}(n,\n)\; z^n \z^{\n} \tau^{-1}\,, \nonumber \\
   &&\hspace*{-1cm} Y_2=\ga_Y(n,\n)\; z^n \z^{\n} \tau^{-1} \,,
   \quad
  \Y_2=\ga_{\Y}(n,\n)\; z^n \z^{\n} \tau^{-1}\,,\quad
   V_2=\ga_V(n,\n)\; z^n \z^{\n} \tau^{-1} \,
\eee
with some constant parameters
$\ga_R(n,\n),\, \ga_W(n,\n),\, ... \sim \gl^{[s]}$, where
$s=1-\half(n+\n)$. The conservation condition means that
$\Phi_2^{n,\n}$ should be ${\cal D}$-closed. The requirement
$E_3^{\cal D} = 0$ modulo terms that do not contribute
to (\ref{3-form}) imposes the following
conditions on the parameters in (\ref{a:t}),
\bee
\label{closed1}
   \hspace*{-2cm}&& 4n\n \ga_R + (n+\n-2)(n\ga_W + \n\ga_{\W})
   + 2n(n-2) \ga_Y  + 2\n(\n-2) \ga_{\Y} = 0 \,, \\
\label{closed2}
   \hspace*{-2cm}&& n\ga_W - \n\ga_{\W}
   + 2(n\ga_Y - \n\ga_{\Y}) = 0 \,, \\
\label{closed3}
   \hspace*{-2cm}&& \ga_V = 0 \,,
\eee
for $n\neq 1$, $\n\neq 1$. For $n=1$ or $\n=1$, $\Phi_2^{n,\n}$
is closed as a consequence of the conditions (\ref{nonzero})
for $E_3^{\cal D}$.

Our problem is to investigate whether there exist the coefficients
$R_1$, $W_1$, ... such that $R_2^{\cal D}$, $W_2^{\cal D}$, ...
have a form (\ref{a:t}). To this end one has to solve the system
(\ref{^R})-(\ref{^bY}) in terms of the formal series
\be
\label{class}
    f(\tau) = \sum_{k=-\infty}^{p<\infty} f_k \tau^k \,.
\ee

Because of the identities (\ref{tF}) there is a freedom in the Fierz
transformations (\ref{delta R})-(\ref{delta bY}) for $\Phi_{1,2}$.
Also one can use the ambiguity in the exact shifts of
$R_1$, $W_1$, ... by any $R_1^{\cal D}$, $W_1^{\cal D}$, ...
(\ref{^phi1}) which do not affect $\Phi_2$ because
${\cal D}^2 =0$. Altogether exact shifts and Fierz transformations
of $\Phi_1$ produce the following equivalence transformations
\bee
\label{d Re}
   \delta R_1 &=& -\dt \chi_1 + \gvep \,,    \\
\label{d We}
   \delta W_1 &=& -\dt \xi_1 + 2i\n \chi_1 - 2(1-i\tau)\gvep \,,  \\
\label{d bWe}
   \delta \W_1 &=& -\dt \bxi_1 - 2in \chi_1 - 2(1+i\tau)\gvep  \,, \\
\label{d Ve}
   \delta V_1 &=& 2(1+\tau^2)\gvep - i(n\xi_1 - \n\bxi_1)  \,, \\
\label{d Ye}
   \delta Y_1 &=& i(\n-1)\xi_1  \,,   \\
\label{d bYe}
   \delta \Y_1 &=& -i(n-1)\bxi_1  \,
\eee
with $\gvep(z,\z,\tau)= -{i\gl\over 2} E_0(z,\z,\tau)$ (\ref{^phi1}).

For a given spin $s$ we consider separately two cases: (i)
with  $n=1$, $\n =1-2s$ or $\n=1$, $n =1-2s$ and (ii)
with  $n < 1$ and $\n < 1$. As shown below, the case (i) corresponds
to the nontrivial physical conserved currents,
whereas the case (ii) describes all possible ``improvements''.

Let us start with the case (i) setting for definiteness $\n=1$.
The case $n=1$ can be considered analogously.
According to (\ref{nonzero}), $Y_2$ is the only coefficient giving
a non-zero contribution to $\Phi^{n,1}_2(y)$. Obviously, a 2-form
with $\n=1$ is invariant under the transformations (\ref{delta Y}).
The only non-trivial equation is (\ref{^Y}). With $Y_2^{\cal D}$
(\ref{a:t}) it takes the form
\be
\label{dY}
   (1+\tau^2)\; {\partial\over \partial\tau}\;Y_1 = -i(1+i\tau)\;Y_1
   + i(2s-1)(1-i\tau)\;Y_1 + i\ga_Y(1-2s,1)\; {z^{1-2s}\z \over \tau}
   + P(\tau) \,,
\ee
where $P(\tau)$ is some polynomial (\ref{arb}).
As shown in Appendix A, the generic solution of (\ref{dY}) is
\bee
\label{Y(tau)}
    Y_1(z,\z,\tau)  = -{i\over 2}\; \ga_Y(1-2s,1) z^{1-2s} \z \;
    (1-i\tau) (1+i\tau)^{2s-1} \ln{(1+\tau^{-2})}
       \nonumber \\
    {}+ \gs \; z^{1-2s}\z \; (1-i\tau) (1+i\tau)^{2s-1} \;
    \ln{\left({1+i\tau^{-1}\over 1-i\tau^{-1}} \right)}+Q(\tau)\,,
\eee
where $\gs$ is an arbitrary constant and $Q(\tau)$ is some inessential
polynomial (\ref{arb}). The logarithms are treated as power series
in $\tau^{-1}$.

At any $\gs$, the solution (\ref{Y(tau)}) is an infinite series
in $\tau^{-1}$, thus corresponding to some pseudolocal 1-form.
Thus, the 2-forms $\Phi^{s}_2(y|x)$ constructed with the
polynomials $Y_2$ at $\n=1$ and with $\Y_2$ at $n=1$ are
${\cal D}$-closed and cannot be represented as $D\Phi^{s}_1(y|x)$
with a spacetime local $\Phi^{s}_1(y|x)$. We therefore argue that
the 2-form $\Phi^{s}_2(y)$ describes a physical conserved
current of spin $s$. The currents (\ref{s-scal})-(\ref{supercurrent})
are reproduced via $Y_2$ (\ref{a:t}) with
\be
   \ga_Y(1-2s,1) = 2^{2s-1}\;(\gl\psi)^{[s]} \,.
\ee
The formula (\ref{Y(tau)}) solves the problem of reformulation of
the physical currents as pseudolocally exact 2-forms.

Let us note that 1- and 2-forms (\ref{int}), (\ref{2-form})
have the following discrete symmetry permutting $C$ and $C'$,
\be
\label{C-C'}
   \Phi\left[C(y),C'(y);\;y\right] = (-)^{\pi(C)\pi(C')+1}\;
   \Phi'\left[C'(iy),C(iy);\;-y\right]
\ee
with $\Phi'(y)$ defined with the parameters
\be
\label{R'R}
   (R',W',\W',V',Y',\Y')(z,\z,\tau) =
   (R,-\W,-W,V,\Y,Y)(-\z,z,\tau)  \,.
\ee
Therefore, the currents generated by $Y_2(n=1-2s,\n=1)$ and
$\Y_2(n=1,\n=1-2s)$ are equivalent by the interchange
$C\leftrightarrow C'$.

Note that the solution (\ref{Y(tau)}) is not unique, containing
an arbitrary parameter $\gs$. Since the transformations
(\ref{d Re})-(\ref{d bYe}) are trivial for $Y_1$ at $\n=1$,
this one-parametric ambiguity cannot be compensated this way.
This means that we have found a pseudolocal 1-form that
is ${\cal D}$-closed but not ${\cal D}$-exact, i.e.,
the cohomology group $H^1 (T,{\cal D})$ is nontrivial.
The physical meaning of this fact is not completely clear to us.
It is however in agreement with the one-parametric ambiguity found
in \cite{PV} for the spin 2 case.

Let us now consider the case of $n<1$, $\n<1$.
Substituting (\ref{a:t}) into the system (\ref{^R})-(\ref{^bY})
we show in appendix B that, if the conditions
(\ref{closed1})-(\ref{closed3}) guaranteeing that $\Phi_2$ is
${\cal D}$-closed are satisfied, then, modulo gauge transformations
(\ref{d Re})-(\ref{d bYe}), its generic solution is
\be
\label{Rt}
    \hspace*{-0.3cm} R_1(z,\z,\tau) = {z^n \z^{\n}\over 4\tau n\n}
    \left[ n\ga_W(n,\n) + \n\ga_{\W}(n,\n)
    - {2n\over \n-1}\, \ga_Y(n,\n)
    - {2\n\over n-1}\, \ga_{\Y}(n,\n) \right]
\ee
($n<0$, $\n<0$),
\be
\label{Wt}
    \hspace*{-0.8cm} W_1(z,\z,\tau) =
    {\ga_{Y}(n,\n)\over \n-1}\; {z^n \z^{\n}\over \tau}
    + \gs (n,\n)\; z^n \z^{\n}\;
    \left[-{i\tau^{-1}\over 1-i\tau^{-1}} + {\n\over 2}\;
    \ln{\left({1+i\tau^{-1}\over 1-i\tau^{-1}} \right)} \right]\,,
\ee
\be
\label{bWt}
    \hspace*{-2.2cm} \W_1(z,\z,\tau) =
    {\ga_{\Y}(n,\n)\over n-1}\; {z^n \z^{\n}\over \tau}
    + \gs (n,\n)\; z^n \z^{\n}\;
    \left[{i\tau^{-1}\over 1+i\tau^{-1}} - {n\over 2}\;
    \ln{\left({1+i\tau^{-1}\over 1-i\tau^{-1}} \right)} \right] \,,
\ee
\be
\label{VYbY}
    \hspace*{-8.8cm} V_1(z,\z,\tau) = Y_1(z,\z,\tau)
    = \Y_1(z,\z,\tau) = 0 \,,
\ee
where $\gs(n,\n)$ are free parameters. At $n = \n = 0$ one should
set $\ga_V(0,0)=0$ and the solution is a pure gauge.

We observe that at $\gs(n,\n) = 0$ the 1-form $\Phi^{n,\n}_1(y)$
leads to a spacetime local expression since $R_1$, $W_1$,
and $\W_1$ (\ref{Rt})-(\ref{bWt}) are linear in $\tau^{-1}$.
Therefore, $\hat{\Phi}^{n,\n}_2(y|x) = D\Phi^{n,\n}_1(y|x)$ with
some local $\Phi^{n,\n}_1(y|x)$. Thus, it is an ``improvement'' of
the physical current 2-form $J(C^2)$ on the r.h.s. of (\ref{Ch-S}),
which can be compensated  by a local field redefinition of
the (higher spin) gauge fields.

Again, the ambiguity related to the parameters $\gs(n,\n)$
is a manifestation of a non-trivial cohomology. We therefore conclude
that $H^1 (T^{n,\n},{\cal D})$ is one-dimensional in each $(n,\n)$
sector. Therefore for a given spin $s=1-\half(n+\n)$,
$\mbox{dim}\; H^1 (T^s, {\cal D}) = 2s+1$, what is of course
the dimension of the spin $s$ representation of the $d3$
Lorentz algebra $o(2,1)$.

Thus we have shown that all local ${\cal D}$-closed forms
$\Phi^{n,\n}_2(y)$ with $n_{\tau}=-1$ are ${\cal D}$-exact
in the class of pseudolocal expansions. The physical conserved
currents are described by $\Phi^{1,1-2s}_2(y)$ or, equivalently,
$\Phi^{1-2s,1}_2(y)$. The ${\cal D}$-closed forms
$\Phi^{n,\n}_2(y)$ with $n,\n \le 0$ are locally ${\cal D}$-exact
and therefore describe various ``improvements''.

\section{Dependence on $\gl$}\label{flat}

Generic $p$-forms given by (\ref{0-form})-(\ref{3-form}) depend on
$\gl$ via the expansions (\ref{Cy}). Such a formulation with
$\gl$-dependent $C(y)$ and $C'(y)$ was convenient for the study of
cohomology since it allowed us to use $\gl$-independent
formulae for the mapping $\cal D$ (\ref{^phi1}),
(\ref{^R})-(\ref{^G}) and the solutions (\ref{Y(tau)}),
(\ref{Rt})-(\ref{bWt}). In the flat limit $\gl\to 0$ the expansion
$C(y)$ becomes meaningless. To investigate what happens in this case
one should use generating functions $\tilde{C}(y)$ (\ref{Cy}).
In this variables, the solutions (\ref{Y(tau)}),
(\ref{Rt})-(\ref{bWt}) acquire explicit dependence on $\gl$.
For example, introducing
\be
   \tilde{\Phi}_{1,2}(\tilde{C},\tilde{C'};\,y) =
   \gl^{-2-\half\pi(C)-\half\pi(C')}\;
   \Phi_{1,2}(C,C';\,\sqrt{\gl}\, y) \,,
\ee
we obtain
\bee
\label{tild}
   \hspace*{-1cm}\tilde{\Phi}_1(\tilde{C},\tilde{C'};\,y)
    &=& h_{\ga\ga} \; {1\over (2\pi)^2}\oint dz  \oint d\z
    \oint \tau^{-2}\,d\tau \int d^2 q \, d^2 \q\;
    \exp\left\{ -{1\over 2\tau} (q_\gga \q^\gga)\right\}
    \nonumber \\
  &&\hspace*{-2cm}
    \times \tilde{C}\left[
    \left.\half(q+\q) - {1\over 2i}(z-\z)y\,,\;\psi\right|x \right]
    \tilde{C}'\left[\left.{1\over 2i}(q-\q) + \half(z+\z)y \,,\;
    \psi\right| x \right]
    \nonumber \\
  &&\hspace*{-2cm}
    {}\times\left\{\tilde{R}_1(z,\z,\tau)\, y^{\ga} y^{\ga}
    +{1\over 2\tau \z} \tilde{W}_1(z,\z,\tau)\, y^{\ga} \q^{\ga}
    +{1\over 2\tau z} \tilde{\W}_1(z,\z,\tau)\, y^{\ga} q^{\ga}
    \right. \nonumber \\
  &&\hspace*{-2cm}\left. {}+{1\over 2\tau^2 \z^2}
    \tilde{Y}_1(z,\z,\tau)\,\q^{\ga} \q^{\ga}
    +{1\over 2\tau^2 z^2} \tilde{\Y}_1(z,\z,\tau)\, q^{\ga} q^{\ga}
    +{1\over 2\tau^2 z\z}
    \tilde{V}_1(z,\z,\tau)\, q^{\ga} \q^{\ga} \right\} \,
\eee
with
\bee
\label{tildR}
    \tilde{R}_1(z,\z,\tau) &=& R_1 (z, \z, \gl\tau)\,, \nonumber\\
    \tilde{W}_1(z,\z,\tau) &=& \gl^{-1}\; W_1(z,\z,\gl\tau) \,,
    \nonumber\\
    \tilde{\W}_1(z,\z,\tau) &=& \gl^{-1}\; \W_1(z,\z,\gl\tau) \,,
    \nonumber\\
    \tilde{Y}_1(z,\z,\tau) &=& \gl^{-2}\; Y_1(z,\z,\gl\tau) \,,
    \nonumber\\
    \tilde{\Y}_1(z,\z,\tau) &=& \gl^{-2}\; \Y_1(z,\z,\gl\tau) \,,
    \nonumber\\
    \tilde{V}_1(z,\z,\tau) &=& \gl^{-2}\; V_1(z,\z,\gl\tau) \,.
\eee
Therefore, in terms of $\tilde{R}$, $\tilde{W}$, ... our solutions
carry an inverse power of $\gl$ together with each power of
$\tau^{-1}$. Equivalently, every spacetime derivation carries
a factor of $\gl^{-1}$. Hence, a representation of physical current
2-forms $\Phi^{s}_2(y)$ as some differentials $D\Phi^{s}_1(y)$
as well as elements of the cohomology group $H^1 (T,{\cal D})$
become meaningless in the flat limit $\gl\to 0$.
In terms of $\tilde{C}(y)$ the formulae (\ref{Y(tau)}), (\ref{Wt}),
(\ref{bWt}) contain the combination $(1\pm i\gl^{-1}\tau^{-1})^{-1}$
to be expanded in powers of $\tau^{-1}$. Viewed as analytic
functions, these formulae have a radius of convergence equal
to $\gl$, so when the AdS radius $\gl^{-1}$ tends to infinity,
the radius of convergence shrinks to zero.

\section*{Conclusion}

In this paper, we construct local conserved currents of an arbitrary
spin in $AdS_3$ built from scalar and spinor fields.
It is shown that they can be treated as ``improvements'' within
the class of infinite power expansions in higher derivatives.
In other words, 2-forms $J$ dual to the physical conserved currents
are shown to be exact in this class, $J=DU$. The 1-forms $U$ are
constructed explicitly what allows us to write down nonlocal field
redefinitions compensating matter sources in the equations of motion
for the Chern-Simons gauge fields of all spins.
The coefficients in the expansion of $U$ in derivatives of the
matter fields contain negative powers of the cosmological constant
(i.e. positive powers of the AdS radius) and therefore do not admit
a flat limit. The existence of $U$ may be related to the holography
in the AdS/CFT correspondence since it indicates that
local current interactions in the AdS space are in a certain sense
trivial and can, up to some surface terms, be compensated by a
field redefinition.

Let us note that our analysis with two independent matter fields
$C$ and $C'$ in (\ref{genform}) covers the case with matter fields
belonging to nontrivial representations of the spin 1 Yang-Mills
group in the extended systems considered in \cite{Eq, PV}.

To analyze the problem systematically for the currents of all
spins we have proposed the formalism of generating functions suitable
for the description of differential forms bilinear in the massless
scalar and spinor fields. It is based on the ``unfolded formulation"
of the dynamical equations as certain covariant constancy
conditions \cite{Unf} and allows us to reformulate the problem
in terms of a cohomology of some differential ${\cal D}$ acting in
the specific auxiliary spaces encoding the full information on the
on--mass--shell matter fields.

Our main result is that the local conserved (i.e. ${\cal D}$-closed)
2-forms of currents, which belong to a non-trivial cohomology within
the class of local expansions, are ${\cal D}$-exact in the class of
pseudolocal expansions (i.e. infinite power series in higher
derivatives). Interestingly enough, we have found that
the cohomology group $H^1(T, {\cal D})$ is nontrivial, implying
nonuniqueness of the solution for $U$ already observed in \cite{PV}
for the case of spin 2. An interesting problems for the future are
to find some group theoretical interpretation of the result that
the dimension of $H^1(T^s, {\cal D})$ in a spin $s$ sector is equal
to the dimension of spin $s$ representation of the $d3$ Lorentz algebra
$o(2,1)$ and to analyze $H^n(T^s, {\cal D})$ for $n\neq 1$.
Since in \cite{PV} it has been shown that there exists a pseudolocal
field redefinition reducing the full nonlinear equations of motion
to the free system we expect that $H^2 (T, {\cal D}) = 0$, but it is
interesting to analyze the problem independently. In particular,
this can shed some light on an appropriate definition of local
functionals in the AdS space as some subspace of the class of
formal pseudolocal expansions.

\bigskip

\noindent
{\bf Note added:}
After this paper had been accepted for publication,
the interesting paper by D.~Anselmi \cite{Ans2} appeared, which contains
explicit expressions for higher spin currents in the flat spacetime
of any dimension, thus generalizing some of the results of \cite{curr}
and the present paper to arbitrary dimensions (in the flat background).

\section*{Acknowledgments}

This research was supported in part by INTAS Grant No.96-0538
and by the RFBR Grants No.96-15-96463, No.99-02-16207, and
No.99-02-17916.
S.~P. acknowledges a partial support from the Landau Scholarship
Foundation, Forschungszentrum J\"ulich.

\setcounter{section}{0}
\def\thesection{Appendix \Alph{section}.}
\def\theequation{\Alph{section}.\arabic{equation}}

\section{Solution at $\n=1$.}\label{n=1}

Consider the following differential equation related to (\ref{dY})
via $x=i\tau$, $k=2s-1$, $\ga = i\ga_Y(n,\n)\; z^{1-2s}\z$, and
$Y = Y_1$,
\be
\label{eq}
   (1-x^2)\; {d\over dx}\;Y = -(1+x)\; Y + k(1-x)\; Y
   + {\ga \over x} + P(x) \,,
\ee
where $k$ and $\ga$ are some constants and $P(x)$ is an arbitrary
polynomial. We have to solve (\ref{eq}) in terms of formal series
(\ref{class}). The essential part of $Y(x)$ contains negative powers
of $x$, i.e. $Y(x)$ is defined modulo arbitrary polynomials.

Rewriting (\ref{eq}) as
\be
\label{eq1}
   (1+x)^{k+1} (1-x)^2  {d\over dx}\;
   \left[(1+x)^{-k} (1-x)^{-1} Y \right] = {\ga \over x} + P(x) \,,
\ee
we solve it in the form
\bee
\label{Y1t}
   Y &=& (1+x)^{k} (1-x) \int^x_c dt\;  (1+t)^{-k-1} (1-t)^{-2}
   \left( {\ga \over t} + P(t) \right) \nonumber\\
   &=& (1+x)^{k} (1-x) \int^x_c dt \left[ {\ga \over t}
   + (1+t)^{-k-1} (1-t)^{-2}  P^\prime (t) \right]
\eee
with the polynomial
$P^\prime (t) = P(t)-\ga t^{-1}\left[(1+t)^{k+1}(1-t)^2-1\right]$.
Using that
\be
   \frac{2}{(1+t)(1-t)} = \frac{1}{(1-t)}+\frac{1}{(1+t)}
\ee
one finds that $P^\prime (t)$ can give a nonpolynomial
contribution to $Y$ only if simple poles in $(1\pm t)$ survive
in the integral (\ref{Y1t}). Equivalently one can set
\be
   P^\prime (t) = (1+t)^{k} (1-t) [\gb (1-t) + \gamma (1+t)]\,.
\ee
Therefore a generic solution of (\ref{eq}) is
\be
   Y(x) = (1-x)(1+x)^k\; [\ga\ln{x} + \gb\ln{(1+x)}
   + \gamma\ln{(1-x)}] \quad mod \quad \mbox{polynomials}\,.
\ee
The restriction to the class (\ref{class}) imposes one restriction
on the parameters $\gb$ and $\gamma$ leading to the final result
\be
\label{gener2}
   Y(x) = -{\ga\over 2}\; (1-x)(1+x)^k\; \ln{(1-x^{-2})}
   + \gs\; (1-x)(1+x)^k \ln{\left({1-x^{-1}\over 1+x^{-1}} \right)}
\ee
with an arbitrary constant $\gs$. Note that the logarithms in
(\ref{gener2}) should be understood as
\be
\label{ln}
    \ln{(1+x^{-1})} = \sum_{m=1}^{\infty} {(-)^{m-1} \over m}
    \; x^{-m} \,.
\ee

\section{Solution at $n<1$, $\n<1$.}\label{n_neq1}

Consider first the case with $n, \n < 0$.
To solve the system (\ref{^R})-(\ref{^bY}) with $R_2^{\cal D}$,
$W_2^{\cal D}$, ... (\ref{a:t}) it is most convenient to gauge fix the
quantities $R_1$, $W_1$, ... in the 1-form $\Phi^{n,\n}_1(y)$
as follows,
\be
\label{gauge}
   V_1 = Y_1 = \Y_1 = 0 \,, \qquad R_1 = {r\over \tau} \,,
\ee
with some $\tau$-independent $r$. Actually, at $n\neq 1$, $\n\neq 1$
one can always achieve (\ref{gauge}) using the transformations
(\ref{d Re})-(\ref{d bYe}). First, one gauges away $Y_1$ and $\Y_1$
with an appropriate choice of the parameters $\xi_1$ and $\bxi_1$.
Taking the parameter $\gvep(\tau)$ in (\ref{d Ve}) in the form
\be
\label{eps}
   \gvep(\tau) = {1\over 1+\tau^2} \;
   \left[ E(\tau^{-1}) + \gvep_0 \tau \right] \,,
\ee
where $E(\tau^{-1})$ is some series in the inverse powers of $\tau$,
and $\gvep_0$ is $\tau$-independent, $V_1$ can be gauged away
by an appropriate choice of $E(\tau^{-1})$. The parameter
$\gvep_0$ remains arbitrary since its contribution to $V_1$ is
proportional to $\tau$ what is equivalent to zero.
{}From (\ref{d Re}) it follows that all terms in $R_1$
can be gauged away with a choice of $\chi(\tau)$
except for the leading term $\tau^{-1}$.
The constant $r$ in the term $r \tau^{-1}$ in $R_1$ can be gauge
fixed by using the ambiguity in $\gvep_0$,
since the corresponding part of $\gvep(\tau)$,
\be
    \gvep_0 \; {\tau\over 1+\tau^2} =
    \gvep_0 \; \sum_{k=0}^{\infty} \; (-)^k \tau^{-(2k+1)} \,,
\ee
contains the term proportional to $\tau^{-1}$. At this stage
however it is convenient to keep $r$ as an arbitrary parameter
to be fixed later. Note that fixing $\gvep_0$ completes gauge
fixing of the transformations (\ref{d Re})-(\ref{d bYe}).

As a result, the system of equations resulting
from (\ref{^R})-(\ref{^bY}) takes the form
\bee
\label{a R}
   \hspace*{-1cm}\ga_R(n,\n)\; {z^n \z^{\n} \over \tau} &=&
   -(n + \n -2)\; {r\over \tau} + {i\over 4}(1+i\tau)\dt  W_1
   - {i\over 4}(1-i\tau)\dt  \W_1   \nonumber \\
   \hspace*{-1cm} &&{}+ {1\over 4}(W_1+\W_1) - \half(nW_1+\n\W_1)
   - \dt \chi_2 \,,  \\
\label{a W}
   \hspace*{-1cm}\ga_W(n,\n)\; {z^n \z^{\n} \over \tau} &=&
   - {i\over 2}\dt
   [(1+\tau^2)W_1] + {3\over 2}(1+i\tau)\;W_1
   + 2\n \; {r\over \tau} \nonumber \\
   \hspace*{-1cm}&& {}+\half (1-i\tau)\n\W_1 - \half (1+i\tau)(\n-1)W_1
   - \dt \xi_2 + 2i\n \chi_2  \,, \\
\label{a bW}
   \hspace*{-1cm}\ga_{\W}(n,\n)\; {z^n \z^{\n} \over \tau} &=&
   {i\over 2}\dt [(1+\tau^2)\W_1]
   +{3\over 2}(1-i\tau)\W_1
   + 2n \; {r\over \tau}  \nonumber \\
   \hspace*{-1cm}&& {} + \half (1+i\tau) nW_1 - \half (1-i\tau)(n-1) \W_1
   - \dt \bxi_2 - 2in \chi_2 \,, \\
\label{a V}
   \hspace*{-1cm}\ga_V(n,\n)\; {z^n \z^{\n} \over \tau} &=&
   \half (1+\tau^2)(nW_1+\n\W_1) - i (n \xi_2 - \n \bxi_2)  \,, \\
\label{a Y}
   \hspace*{-1cm}\ga_Y(n,\n)\; {z^n \z^{\n} \over \tau} &=&
   \half (1+\tau^2)(\n-1)W_1 + i(\n-1) \xi_2  \,,  \\
\label{a bY}
   \hspace*{-1cm}\ga_{\Y}(n,\n)\; {z^n \z^{\n} \over \tau} &=&
   \half (1+\tau^2)(n-1)\W_1 - i(n-1) \bxi_2  \,,
\eee
where all the equalities are treated modulo polynomials (\ref{arb})
and the variables $\chi_2$, $\xi_2$, and $\bxi_2$
account for the ambiguity modulo the Fierz transformations of
$R^{\cal D}_2$, $W^{\cal D}_2$, ... .

Introducing the new variables
\be
\label{newvar}
   X = nW_1 - \n\W_1 \,,\qquad X^+ = nW_1 + \n\W_1 \,,
\ee
and expressing the Fierz parameters $\chi_2$,
$\xi_2$, and $\bxi_2$ in terms of the rest variables,
one reduces (\ref{a R})-(\ref{a bY})  to
\be
\label{X+}
   (1+\tau^2)\; X^+ = {A\over \tau} + P_1(\tau)\,,
\ee
\be
\label{X}
   (1+\tau^2)\; {\partial\over \partial\tau}\; X =
   -2i X^+ - {i(4n\n\; r-G)\over \tau}
   - {B\over \tau^2} + P_2(\tau)  \,,
\ee
\bee
\label{consis}
   &\hspace*{-2cm}& 4n\n\; {\ga_R\over \tau}
   + 4n\n (n+\n-2)\; {r\over \tau}
   - \left[ \tau(n-\n) + i(n+\n-2) \right] \dt X \nonumber \\
   &\hspace*{-2cm}& {} + \left[ i(n-\n) + \tau(n+\n-2) \right] \dt X^+
   + 2(n\n-1)X^+ + {iK \over \tau^2} + {A-L \over \tau^3}
   = P_3(\tau) \,,
\eee
where
\bee
\label{notat}
   A &=& \left( {n\over \n-1}\; \ga_Y
     + {\n\over n-1}\; \ga_{\Y} + \ga_V \right)\; z^n \z^{\n}  \,,
     \nonumber \\
   B &=& \left( {n\over \n-1}\; \ga_Y
     - {\n\over n-1}\; \ga_{\Y} \right)\; z^n \z^{\n} \,,
     \nonumber \\
   G &=& \left( n\ga_W + \n\ga_{\W} \right)\; z^n \z^{\n} \,,
     \nonumber \\
   K &=& \left( n\ga_W - \n\ga_{\W} \right)\; z^n \z^{\n} \,,
     \nonumber \\
   L &=& \left( {n\over \n-1}\; \ga_Y
     + {\n\over n-1}\; \ga_{\Y} - \ga_V \right)\; z^n \z^{\n}  \,,
\eee
and $P_1(\tau)$, $P_2(\tau)$, and $P_3(\tau)$ are arbitrary
polynomials.

Using that any polynomial $P(\tau)$ can be rewritten as
\be
   P(\tau)= a + b \tau + (1+\tau ^2 ) p(\tau)
\ee
with some polynomial $p(\tau)$, one obtains the general solution
of (\ref{X+}),
\be
\label{sol.X+}
   X^+(\tau) = {A\over \tau} + \gs^+_1 \; z^n \z^{\n} \;
     {1\over 1+\tau^2}
     + \gs^+_2 \; z^n \z^{\n} \;{\tau\over 1+\tau^2} + p(\tau) \,,
\ee
where $\gs^+_1$ and $\gs^+_2$ are arbitrary constants.

Inserting (\ref{sol.X+}) into (\ref{X}) and solving it for $\dt X$
analogously to (\ref{X+}), we get
\bee
\label{eq1X}
   {\partial\over \partial\tau}\; X &=&
   - {i[4n\n\;r - (G-2A)]\over \tau (1+\tau^2)} - {B\over \tau^2}
   -2i \gs^+_1 \; z^n \z^{\n} \; {1\over (1+\tau^2)^2}
   -2i \gs^+_2 \; z^n \z^{\n} \; {\tau\over (1+\tau^2)^2}
   \nonumber \\
   && {} + \gs_1 \; z^n \z^{\n} \; {1\over 1+\tau^2}
   + \gs_2 \; z^n \z^{\n} \; {\tau\over 1+\tau^2}  \,,
\eee
where $\gs_1$ and $\gs_2$ are arbitrary constants.
Now it is convenient to set
\be
\label{rgauge}
   r = {G-2A \over 4n\n}\,.
\ee
Also one should set $\gs_2 = 0$ in (\ref{eq1X}) since the term
$\tau (1+\tau^2)^{-1}$ is not integrable in the form (\ref{class}).
As a result, we arrive at the differential equation
\be
\label{eqX}
   {\partial\over \partial\tau}\; X = - {B\over \tau^2}
   -2i \gs^+_1 \; z^n \z^{\n} \; {1\over (1+\tau^2)^2}
   -2i \gs^+_2 \; z^n \z^{\n} \; {\tau\over (1+\tau^2)^2}
   + \gs_1 \; z^n \z^{\n} \; {1\over 1+\tau^2}  \,,
\ee
and therefore, modulo polynomials,
\bee
\label{solX}
   X(\tau) &=& {B\over \tau}
   -i \gs^+_1 \; z^n \z^{\n} \; \left[{\tau^{-1}\over 1+\tau^{-2}}
   + {i\over 2}\; \ln{\left({1+i\tau^{-1}\over 1-i\tau^{-1}} \right)}
   \right] + i \gs^+_2 \; z^n \z^{\n} \; {\tau^{-2}\over 1+\tau^{-2}}
   \nonumber \\
   &&{} + \gs_1 \; z^n \z^{\n} \;
   {i\over 2}\; \ln{\left({1+i\tau^{-1}\over 1-i\tau^{-1}}
   \right)} \,.
\eee

The equation (\ref{consis}) is equivalent to the conditions
\be
\label{par1}
   (2n\n-n-\n)\;\gs^+_2 - (n-\n)\; \gs_1 = 0 \,,
\ee
\be
\label{par2}
   2(n\n-n-\n+1)\; \gs^+_1 + i(n-\n)\; \gs^+_2
   - i(n+\n-2)\; \gs_1 = 0 \,
\ee
on the parameters $\gs_1$, $\gs^+_1$ and $\gs^+_2$ provided that
the conditions (\ref{closed1})-(\ref{closed3}) guaranteeing that
$\Phi^{n,\n}_2(y)$ is ${\cal D}$-closed are satisfied.

Using (\ref{gauge}), (\ref{newvar}), (\ref{notat}), (\ref{sol.X+}),
(\ref{rgauge}), (\ref{solX})-(\ref{par2}), one finds the general
solution of the system (\ref{a R})-(\ref{a bY})
in the form (\ref{Rt})-(\ref{VYbY}), where $\gs = i\gs_1$.

Let us turn to the case with $\n=0$, $n < 0$
(the case with $n=0$, $\n < 0$ can be considered analogously).
According to (\ref{nonzero}) the only parameters giving a non-zero
contribution to $\Phi^{n,\n}_{1,2}(y)$ are $W_{1,2}$, $V_{1,2}$,
and $Y_{1,2}$, so that the nontrivial equations in the system
(\ref{a R})-(\ref{a bY}) are (\ref{a W}), (\ref{a V}),
and (\ref{a Y}). Analogously to the case considered above,
one gauge fixes
\be
\label{gauge2}
   V_1 = Y_1 = 0
\ee
and obtains the solution
\be
\label{Wt2}
    \hspace*{-7cm} W_1(z,\z,\tau) =
    -\ga_{Y}(n,0)\; {z^n \over \tau}
    + \gs (n,0)\; z^n \;  {\tau^{-1}\over 1-i\tau^{-1}} \,.
\ee

The rest case $n,\n=0$ is trivial. Indeed, in this case the only
contributing parameters are $\ga_V(0,0)$ and $V_1$. The parameter
$\ga_V(0,0)$ vanish by the equation (\ref{a V}), while $V_1$ can be
gauged away by the transformation (\ref{d Ve}).

\end{document}